\documentclass[12pt]{article}
\usepackage[centertags]{amsmath}
\usepackage{amssymb}
\usepackage{amsthm}
\usepackage{enumerate}
\usepackage{url}
\usepackage{graphicx}

\usepackage{graphicx}
\usepackage[all, knot]{xy}
\usepackage{amsfonts,amssymb}
\usepackage{latexsym}
\usepackage{amsmath}
\usepackage[all]{xy}
\usepackage{stmaryrd}

\usepackage[english]{babel}

\usepackage[small,nohug,heads=vee]{diagrams}
\diagramstyle[labelstyle=\scriptstyle]

\usepackage{color}
\usepackage[normalem]{ulem}
\usepackage{url}

\makeatletter
  \def\tagform@#1{\maketag@@@{(#1)\@@italiccorr}}
\makeatother

\newcommand {\cE}{\mbox{${\mathcal E}$}}

\newcommand {\cM}{\mbox{${\mathcal M}$}}

\newcommand{\R}{\mathbb{R}}

\newcommand{\NN}{\mathbb{N}}

\begin{document}

\title{Gravity in the Smallest}
\author{Michael Heller \\
Copernicus Center for Interdisciplinary Studies\\
ul. Szczepa\'nska 1/5, 31-011 Cracow, Poland\\
\and 
Jerzy Kr\'ol \\
Institute of Physics, University of Silesia, \\
ul. Uniwersytecka 
4, 40-007 Katowice, Poland}

\date{\today}
\maketitle

\begin{abstract}
Synthetic Differential Geometry (SDG) is a categorical version of differential geometry based on enriching the real line with infinitesimals and weakening of classical logic to intuitionistic logic. We show that SDG provides an effective mathematical tool to formulate general relativity in infinitesimally small domains. Such a domain is modelled by a monad around a point $x$ of a manifold $M$, defined as a collection of points in $M$ that differ from $x$ by an infinitesimal value. Monads have rich enough matematical structure to allow for the existence of all geomeric quantities necesary to construct general relativity ``in the smallest''. We focus on connection and curvature. We also comment on the covariance principle and the equivalence principle in this context. Identification of monads with what happens ``beneath the Planck threshold'' could open new possibilities in our search for quantum gravity theory.

\end{abstract}

\section{Introduction}
There is a growing feeling that the present rather slow progress in approaching the correct theory of quantum gravity is due to our misunderstanding of some of fundamental physical principles, when they are applied to very small scales (see for instance \cite{Baez,tovariance}). After all, there is no \textit{a priori} guarantee that what well organises our knowledge on macroscopic or near-macroscopic level, behaves equally well when we are going down some thirty or more orders of magnitude\footnote{Planck distance $l_P = \sqrt{\hbar G/c^3}$ is of order of $10^{-35}$ m.} along the distance ladder. The authors, quoted above, see a remedy to this situation in a careful examination of our principles with the help of methods provided by category theory. The work has been taken over, first of all in the field of quantum mechanics \cite{AC,DoringIsham,tovariance,bohrif2,Landsman16}, but also as far as general relativity is concerned \cite{Grin96,GutsGrin96,GutsZvy,tovariance}, and in general physics as well \cite{Baez,BaezDolan}. Our approach in this paper goes along these lines, but we adopt slightly different strategy. We are interested in discussing some fundamental principles of general relativity when an attempt is made to apply them to extremely small distances which usually are supposed to be controlled by quantum gravity; in other words, we want to explore some aspects of gravity ``in the smallest''.

Mathematical tools to undertake this enterprice are almost ready. Within category theory there exists a version of differential geometry called Synthetic Differential Geometry (SDG for brevity) \cite{Kock1984,Kock09,Kostecki09,MoerRey}. It is based on enriching the real line $\R $ with infinitesimals (different than those met in non-standard analysis) which is enforced by placing differential geometry into a suitable category. This is done not without a price: to ensure the consistency of the procedure, the underlying logic has to be weakened -- the law of excluded middle must be blocked, and non-constructive proofs forbidden. It is the existence of infinitesimals that makes it possible to penetrate ``infinitesimal portions'' of a manifold which are inaccessible when standard tools are employed. In the present paper, we use well known results of SDG (mainly as it is presented in \cite{Kock09}), our original idea being to shrink the analysis to ``infinitesimal neighbourhoods'' (to be defined below). Physically, we should think of such neighbourhoods as being beneath the Planck scale.

Roughly speaking, there are two versions of SDG: one, closely following traditional pattern, is based on the tangent bundle over a manifold $M$, and the other, called by Kock combinatorial approach \cite{Kock09}, exploiting a binary relation between points of $M$, two points being interrelated if the ``distance'' between them is infinitesimal. It is the second approach that  provides the formalism suitable for investigating geometry ``in the smallest''. Analyses carried out in the present paper are based entirely on the second approach. 

SDG is usually developed in an axiomatic or quasi-axiomatic method \cite{Kock1984,Kock09}, but natural models for SDG are provided by some categories, especially those in which the category of manifolds (and smooth functions between them as morphisms) can be embedded. Such categories were studied by Dubug \cite{Dubug}, called by him ``well adapted'' categories, and especially extensively by Moerdijk and Reyes \cite{MoerRey} under the name of ``smooth categories''. Which of possible models could find their application to physics is an open problem, possibly pregnant with consequences.

The existence of infinitesimals has another consequence which could be important as far as physical applications are concerned. As we shall see in section 2, owing to their existence, operations based on the traditional concept of limit are effectively replaced by purely algebraic operations. In section 2, we also briefly discuss how logic is involved in the strategy of infinitesimals, and introduce concepts that are primary tools of our further analysis: different kinds of infinitesimals and the $k$th order neighbouring relation. In section 3, this relation is generalised to the manifold context, and serves to define the notions of a $k$-monad and a ``$k$th neighbourhood of the diagonal''. 

In section 4, we introduce affine connection and parallel transport in terms of it. It turns out that this connection is flat. Everything is discussed within ``infinitesimally small neighbourhoods''. The same regards section 5, in which a psedo-Riemannian metric is defined  that, after imposing some additional conditions, can be turned into a positively defined Riemannian metric. We also introduce  a quasi-metric function which has an interesting interpretation.

In section 6, we employ the above geometric machinery to investigate the interaction between connection and curvature on infinitesimal neighbourhoods with those on macro-regions. This is directly related to Einstein's weak equivalence principle which asserts that gravity can locally be always eliminated. Our result shows that the mathematical content of this principle  has its origin in purely geometric properties of infinitesimal neighbourhoods.

Some comments related to principles on which general relativity is based (covariance principe, weak and strong equivalence principles), and how they operate on ``infinitesimal neighbourhoods'', are collected in section 7. If we curageously identify ``infinitesimal neighbourhoods'' with ``beneath the Planck threshold'', this analysis could be of some value for our search of quantum gravity theory. It also can have some merit for better understanding the standard approach to general relativity.

\section{Infinitesimals and Logic}
In SDG the real line $\R $ is enriched with infinitesimals; the enriched $\R $ will be denoted by $R$. To the most fundamental infinitesimals belong the so-called nilpotent infinitesimals: they are so small, but not necessarily equal to zero, that their square vanishes; we denote them by $D$,
\[
D:=\{x \in \R | x^2 =0\}.
\]
The consequences of the existence of $D$ are radical. For instance, let us compute the derivative of the function $f(x) = x^2$ at $x = c$. For $d \in D$ we have
\[
f(c+d) = (c+d)^2 = c^2+2cd + d^2 = c^2 + (2c)d.
\]
The linear part of this expression can be identified with the 
derivative of $f(x)$ at $c$, $f'(c) = 2c$. This example suggests the following rule

For any $g: D \rightarrow R$, there exists a unique $b \in 
R$ such that
\begin{equation}
\label{Axiom}
\forall d \in D, g(d) = g(0) + d \cdot b.
\end{equation}
In SDG this rule is elevated to the rank of an axiom. It implies that the graph of $g$ coincides with a fragment of the straight line through $(0, g(0))$ and the slope $b$. The derivative of any function $f: R \rightarrow R$ at $c$ can be defined to be $f'(c) = b$. In fact, in SDG all functions are differentiable (for a friendly introduction see \cite{Pizza}). It is obvious that this must have far-reaching consequences for the space-time geometry in the smallest.

Let us consider the function 

\[
g(d) =
\left\{
\begin{array}{cc}
1 & \mbox{if $d \neq 0$} \\
0 & \mbox{if $d = 0$}
\end{array}
\right.
\]

By the law of excluded middle and the fact that $D \neq \{0 \}$, there exists $d_0 \neq 0$ in $D$ which, on the strength of equation (\ref{Axiom}), gives
\[
g(d_0) = g(0) + d \cdot b.
\]
This leads to $1 = g(d_0) = d \cdot b$, and after squaring we get $1 = 0$, the evident contradiction.

The only way to save our reasoning is to block the law of excluded middle. And indeed, the SDG works on the basis of weakening classical logic to the intuitionistic logic, in which the law of excluded middle does not hold. Of course, one cannot change logic at will. A modified logic requires a correct structural environment that would not only justify but also enforce the correct modification of logic. It is provided by suitable categories.

In SDG, besides $D$, there are various kinds of infinitesimals\footnote{In general, infinitesimal objects are identified with the spectrum of a Weil algebra; see Appendix A.1.}; in the following, we make use only of a few of them:
\[
D_k=\{ x\in R|x^{k+1}=0 \}, \; k=1,2,3,..., 
\] 
\[ 
D(n)=\{ (x_1,...,x_n)\in R^n|x_ix_j=0,\; \forall i,j=1,2,3,...,n \}, 
\] 
\[
 D_k(n)=\{(x_1,...,x_n)\in R^n|{\rm \, the\, product\, of \, any}\,  
k+1\, {\rm of} \, x_i\, {\rm is}\, 0 \}, 
\] 
and finally, 
\[
 (D_{\infty})^n=\bigcup_{k=1}^{\infty}D_k(n).
\]

Analogously, we can define $D(V)$ and $D_k(V)$ for any finite dimensional vector space.

In what follows, our important tool is the ``$k$th order neighbouring relation'', defined as
\[
u \sim_k v \Leftrightarrow  u - v \in D_k(V).
\]
This relation is reflexive and symmetric, but it is not transitive; instead we have: if $u\sim_kv$ and $v\sim_l w$ then $u \sim_{k+l} w$.

\section{Manifolds in a Categorical Context}
We assume that everything in this section happens in a ``suitable'' category \cE \ such that a space $M$ belongs to its objects. ``Suitable'' means a category equipped, among others with a commutative ring object $R$ and usually a topos.

By an $n$-dimensional manifold we understand a space $M$ such that there exists a \textit{family} $\{U_i | i \in I\}$ of spaces that is equipped with \textit{open} inclusions $U_i \to M$ and $U_i \to R^n$, and the family $U_i \to M$ is supposed to be jointly surjective with respect to $M$. This description crucially depends on the meaning of two terms: ``family'' and ``open''. The meaning of the ``family of spaces'' is taken in the external sense: the index set $I$ in the formula $U_i \to M$ is not an object in \cE , but rather an external discrete set. In this intuitive approach, we simply assume that in \cE \ there is a subclass $\mathcal{R}$ of arrows regarded as ``open inclusions''. Formally, they are defined as monic \'etale maps (see Appendix A.2). Alternatively, we could replace the manifold concept, as described above, with the concept of a formal $n$-dimensional manifold; its definition is given in Appendix A.2.

The families of open inclusions $U_i \to M$ and $U_i \to R^n$ are said to form an atlas on $M$, and the single maps $U_i \to R^n$ are said to be coordinate charts or coordinate systems of this atlas.

We are interested in the ``smallest neighbourhoods'' of an $n$-dimensional manifold $M$. A good tool to investigate such neighbourhoods is the neighbourhood relation $\sim_k$, which we now generalise to the manifold context. Let $x, y \in M$ and $k$ a non-negative natural number, the relation $x \sim_k y$ holds iff there exists a coordinate chart $f: U \to M$ such that $U\subseteq R^n$ is open, and in $U$ we have $f(x) \sim_k f(y)$. If $k=1$, we simply write $x\sim y$.

In the above description of an $n$-dimensional manifold $M$, we could say that $M$ is modelled on $R^n$. Analogously, in many situations $M$ can be modelled by an abstract $n$-dimensional vector space $V$.\footnote{In SET, any manifold $M$ can locally be considered as an open subset of a finite dimensional vector space $V$ the dimensionality of which coincides with that of $M$. In $\cE $, a model vector space $V$ is over the ring $R$, thus $R^n \simeq V$ for some $n \in \mathbb{N}$.}

Now, we are going to define a few concepts indispensable for our further analysis.

Let $x \in M$. The $k$-monad around $x$ is defined to be
\[
\cM_k(x) := \{y\in M| x\sim_k y\} \subseteq M.
\]
If $k=1$, we simply write $\cM(x)$. Since the relation $\sim_k $ is reflexive, $x \in \cM_k(x)$, and since $\sim_k$ is symmetric, we have $y\in \cM_k(x) \Leftrightarrow x\in \cM_k(y)$. Any map between manifolds $M$ and $N$, $f: M\to N$, preserves the relation $\sim_k$: if $x\sim_k y$ then $f(x) \sim_k f(y)$.

The ``$k$th neighbourhood of the diagonal'', $M_{(k)} \subseteq M \times M$, is defined to be
\[
M_{(k)} := \{(x,y) \in M \times M|x \sim_k y\}.
\]
If $V$ is an $n$-dimensional vector space, there is a canonical isomorphism
\[
M_{(k)}\cong M \times D_k(V)
\]
given by
\[
(x,y) \mapsto (x, y-x).
\]

A triple of points $x,y,z \in M$ is said to form an infinitesimal 2-whisker at $x$ if $x\sim y$ and $x\sim z$; if, additionally, $y \sim z$ it is said to form an infinitesimal 2-simplex at $x$.

Let $M$ be a manifold and $x \in M$. A bijection $k_x: D(n) \to \cM(x)$ that takes $0$ to $x$ is said to be a frame at $x$. If such a frame is given for every $x \in M$, we say that there is a framing of $M$. The set of all frames at all points of $M$ is said to be a frame bundle of $M$ (it is a principal fibre bundle), and framing may be defined as its cross section.

The inverse of $k_x$
\[
k_x^{-1} = c_x: \cM (x) \to D(n)
\]
defines coordinates for the neighbourhood of $x$: coordinates of $y\in \cM(x)$ are $c_x(y) \in D(n) \subseteq R^n$. We should notice that $c_x(x) = 0$.

Let us consider a frame $k_x: D(n) \to \cM(x)$. It is interesting to notice that if $M \subseteq V$, there exists the \textit{canonical} framing
\[
k_x(d) = x + d,
\] 
and the \textit{canonical} coordinates for $y \in \cM(x)$ in this frame are
\[
c_x(y) = y - x \;\; \mathrm{for} \; y \sim x.
\]
% czyli wszystkie układy wspólrzędnych różnią się o infinitesimalę.

\section{Connection}
Theory of connection in this formalism was elaborated by Kock \cite{Kock1984,Kock1998}, \cite[chapt. 2.3]{Kock09}. An affine connection on a manifold $M$ is defined to be a rule $\lambda $ that to any infinitesimal 2-whisker in $M$ associates a point $\lambda (x,y,z)$ such that
\begin{equation}
\label{A}
\lambda(x,y,x)=y
\end{equation} 
\begin{equation}
\label{B}
\lambda(x,x,z)= z.
\end{equation}
If we remember that an infinitesimal whisker is a triple of points $(x,y,z)$ such that $x\sim y$ and $x\sim z$, then the above definition amounts to closing the whisker into an infinitesimal parallelogram. $\lambda $ also satisfies
\begin{equation}
\label{C}
\lambda(y,x,\lambda({x,y,z})) = z.
\end{equation}

A connection is said to be torsion free if it is symmetric with respect to the last two arguments, i.e. if $\lambda(x,y,z) = \lambda(x,z,y)$.

For $x\sim y$, $\lambda(x,y,-)$ defines a map
\[
\lambda(x,y,-): \cM(x) \to \cM(y)
\]
which sends $x$ to $y$ (by eq. (\ref{A}), is identity if $x=y$ (by eq. (\ref{B})), and has an inverse $\lambda(y,x,-)$ (by eq. (\ref{C})). Therefore, it can be regarded as a ``parallel transport'' from $x$ to $y$. Let us introduce the following abbreviation: $\nabla(y,x) = \lambda(x,y,-)$.

Let us now consider a triple of points $x_1, x_2, x_3 \in M$ with $x_1\sim x_2, \, x_2\sim x_3, \, x_3\sim x_1$, i.e. these points form a ``triangle''. Let us consider the following diagram

\begin{diagram}
\cM(x_1)      &\rTo^{\nabla(x_2,x_1)}       &\cM(x_2) \\
              &\luTo_{\nabla(x_1,x_3)}      &\dTo_{\nabla(x_3,x_2)}\\
              &                                &\cM(x_3)
\end{diagram}

A connection $\lambda $ is defined to be curvature free (or flat) if the composite map $\cM(x_1) \to \cM(x_1)$ for any such triangle is the identity map on $\cM(x_1)$ or, equivalently, if for any infinitesimal 2-simplex $(x_1,x_2,x_3)$ in $M$, one has \cite[p. 59]{Kock09}
\[
\nabla(x_3,x_1) = \nabla(x_3,x_2) \circ \nabla(x_2,x_1).
\]

We choose $\cM(x), \, x \in M$, and a frame $k_x: D(R^n) \to \cM(x)$ with the corresponding coordinates $c_x =k_x^{-1}$. We define an affine connection on $M$
\[
\lambda(x_1,x_2,x_3) := k_{x_2}(k_{x_1}^{-1}(x_3)) = c_{x_2}^{-1}(c_{x_1}(x_3)),
\]
which can be read as ``the point $\lambda(x_1,x_2,x_3)$ is that point which in the coordinate system at $x_2$ has the same coordinates as $x_3$ does in the coordinate system at $x_1$'' \cite[p. 61]{Kock09}.

This connection is flat. Indeed, for an infinitesimal 2-simplex $(x_1,x_2,x_3)$ in $M$ we have
\[
\nabla(x_3,z_2)\circ \nabla(z_2,z_1) = k_{x_3} \circ k_{x_2}^{-1} \circ k_{x_2} \circ k_{x_1}^{-1} = k_{x_3} \circ k_{x_1}^{-1} = \nabla(x_3,x_1).
\]

Moreover, if this connection comes about from the framing $k_x$, for every $x \in M$, on $M$, then it is flat on $M$ \cite[Proposition 3.7.3]{Kock09}.

\section{Metric Structure}

A quadratic differential form on a manifold $M$ is defined to be a map $g: M_{(2)}\to R$ vanishing on $M_{(1)} \subseteq M_{(2)}$ (the latter condition generalises the classical fact that $g(x,x) = 0$). It can be demonstrated that $g$ is symmetric, i.e. $g(x,y)=g(y,x)$ \cite[Proposition 8.1.2]{Kock09}. As a ``canonical example'' we have
\[
g(x,y) = \sum_{i=1}^n(y_i-x_i)^2
\]
which looks as representing the square distance between $x$ and $y$; however, the concept of distance itself can hardly be made meaningful in this context. We should remember that $x\sim_k y$ for any $k\geq 1$.

If $V$ is a finite dimensional vector space, $g$ can be identified with the map
\[M \times D_{(2)}(V) \to R
\]
vanishing on $M \times D_{(1)}(V)$. The identification is provided by $(x,y) \mapsto (x, y-x)$, and $g$ can be written as
\[
G(x;-,-): V \times V \to R,
\]
given by
\[
g(x,y) = G(x;\; y-x,y-x)
\]
which is bilinear and symmetric \cite[p. 251]{Kock09}.

Let $V^*$ denote the vector space of linear maps $V \to R$. A bilinear form $B: V \times V \to R$ induces a linear map $\hat{B}: V \to B^*$, $\hat{B}(u)(v) := B(u,v)$. A bilinear form $B: V \times V \to R$ is nondegenerate if $\hat{B}: V \to V^*$ is an isomorphism. Now we are ready to define a metric on a manifold.

A pseudo-Riemannian metric on a manifold $M$ is defined to be a quadratic differential form $g$ on $M$ if for some (and consequently, for every) local coordinate system around $x \in M$, with $M$ an open subset of a finite dimensional vector space $V$, the map
\[
G(x;-,-): V \times V \to R,
\]
given by
\[
G(x; v,w) = \frac{1}{2}(g(x,v+w) - g(x,x+v) - g(x,x+w)),
\]
$v,w \in V$, is nondegenerate. $G$ is called the metric tensor of the metric $g$ relative to the ``coordinate system'' $V$ (see \cite[p. 253]{Kock09}).

Let $g$ be a quadratic differential form on a manifold $M$, and $\lambda $ an affine connection on $M$. The connection $\lambda$ is said to be compatible with $g$ if the transport
\[
\nabla(y,x): \cM_1(x) \to \cM_1(y)
\]
preserves $g$. If $g$ is a pseudo-Riemannian metric on a manifold $M$, there exists a unique symmetric affine connection $\lambda $ on $M$ that is compatible with $g$ \cite[Theorem 8.1.4]{Kock09}.

In order to have a positive definite metric some additional conditions must be satisfied.

A vector $x = (x_1, \dots , x_n)$ is said to be proper if at least one of $x_i$s is invertible.

An element $b$ of $R$ is said to have a square root if there is $c \in R$ such that $c^2 = b$.

An inner product space is a finite dimensional vector space $V$, equipped with a symmetric bilinear form $\left\langle -,-\right\rangle: V \times V \to R$ such that for every proper vector $v \in V$, $\left\langle v,v \right\rangle$ is invertible and has a square root.

And finally, a symmetric bilinear form $V \times V \to R$, changing $V$ into an inner product space, is said to be positive definite. If on $R$ there is defined the concept of a strict positivity, such that all positive elements in $R$ are invertible and have square roots, then the concept of inner product space in this definition can be replaced by demand that, for all $a \in V$, $a \in R$, $\left\langle a, a\right\rangle $ is strictly positive.

Let us finally mention a concept that makes sense in an ``infinitesimal neighbourhood'' and is quite close to the concept of metric. Let $M$ be a manifold and $x,y,z \in M$. The following conditions are satisfied
\begin{enumerate}
\item
$x \sim_0y$ iff $x=y$ (reflexivity).
\item
$x\sim_ky$ implies $y\sim_lx$ if $k\leq l$ (symmetry),
\item
$x\sim_ky$ and $y\sim_lz$ implies $x\sim_{k+l}z$ (quasi-trangle formula).
\end{enumerate}
We can define a ``quasi-distance'' function
\[
\mathrm{dist}(x,y) \leq k \;\; \mathrm{if} \;\; x\sim_ky.
\]
This function is almost like a discrete metric with values in $\NN $.

\section{The Principle of Equivalence in a Topos}
In this section we apply the above mathematical machinery to study some aspects of general relativity ``in the smallest''. Usually, general relativity, just as all other macroscopic theories, is considered in the set theoretical environment (in the category SET of sets as objects and functions between sets as morphisms). To make use of infinitesimals, we must change from the category SET to a ``suitable category'' $\cE $, just as in the beginning of section 3. We shall assume that $\cE $ is a topos.

We propose the following intuitive image. On macroscopic scales we have the usual general relativity (in the SET environment); when we go down to smaller and smaller scales, finally reaching the Planck thershold, we find ourselves in the $\cE $ environment.\footnote{How to mathematically describe the process of transition between categories requires a separate study which is under way.} This suggests a careful looking into some basing principles of general relativity.

In general, there is plenty of both flat and non-flat connections defined on monads $\cM(y) $ and macro regions as well. There is, however, a strong interaction of curvature on infinitesimal neighbourhoods with that on macro-regions; this interaction is clarified by the following proposition.

\textbf{Proposition.}
\textit{Let a manifold $M \subseteq R^n$ be an object of $\cE $, and $U_x, \, x\in M$, open in $M$. There exists a flat connection $\lambda $ on $U_x$ if and only if the connection $\lambda $ locally (on every $\cM(y),\, y\in U_x$) comes about from a framing.}

The proof follows from combining Propositions 2.4.1 and 3.7.3 of \cite{Kock09}; we give only an outline of the reasoning. Let an affine connection $\lambda $ be given by coordinates (framings) on every $\cM(y),\, y \in U_x$, i.e. $\lambda(x,y,z)=k_y\circ  k_x^{-1}$. Directly from Proposition 2.4.1 of \cite{Kock09}, and from the construction as at the end of section 4 above, it follows that $\lambda$ is the (unique) flat affine connection on the entire $U_x$ and it comes about from the framings $k_y$. 

On the other hand, let a flat connection $\lambda $ exists on $U_x, \, x\in M$, i.e. in terms of infinitesimal transport on every infinitesimal 2-simplex $(p,y,z)\subset U_x$ one has
\[ \nabla(p,z)=\nabla(p,y)\circ \nabla(y,z) .\] To show that such $\lambda$ comes about locally from some framing $k$ we define locally an auxiliary framing $h$ on $U_x$ by
\[
h_y: D(n) \to \cM(y), \; \; y \in U_x.
\]
Then we consider a map  
\[
g: U \to GL(R^n), U\subset U_x
\]
where $GL(R^n)$ is a group of linear automorphisms of $R^n$ which can, by the axioms of SDG, be identified with invertible and zero-preserving maps $D(n) \to D(n)$. We take the inverse of such a map
\[
g(y)^{-1}: D(n) \to D(n),
\]
and, with the help of it, define a new framing
\[
k_y = h_y \circ g(y)^{-1},
\]
from which the connection $\lambda $ locally comes about, i.e. $\lambda=k_y\circ k^{-1}_z$ on every 2-simplex $(p,y,z)\subset U_x$. Details of this construction are given in the proof of Proposition 3.7.3 of \cite{Kock09}. $\Box $

Let us notice that if the connection on monads $\cM(y), y \in U_x$, is associated with the canonical framing $k_y: D(n) \to \cM(y), \, d \mapsto y + d$, then it is flat on $U_x$.

Since the existence of a flat connection on $U_x, \, x\in M$, with $M$ interpreted as a space-time manifold, expresses the mathematical content of Einstein's Weak Equivalence Principle, the above proposition shows, in fact, the roots of this Principle: its origin is purely geometric, provided the geometry is enriched with infinitesimals.

\section{An Interplay of Concepts}
In looking for the theory of quantum gravity, we often (implicitly) assume that the same fundamental principles, that led Einstein to general theory of relativity, either remain in force also at the quantum gravity level or should be suitably generalised. Does mathematics underlying the categorical approach to geometry confirm this view?

Two of such principles are of special interest in the present context: the principle of covariance and the principle of  equivalence (nice analysis of these principles in the standard approach can be found in \cite[chapt. 3]{EllisAll}). They were analysed by Heunen, Landsman and Spitters, under the assumption that the unification of gravity and quanta could require theory of topoi as its logical environment \cite{tovariance}. According to these authors, in the topos environment the principle of general covariance should assume the form of what they call the principle of general tovariance; it states that ``any mathematical structure appearing in the laws of physics must be definable in an arbitrary topos (with natural numbers object) and must be preserved under geometric morphisms.'' Roughly speaking, geometric morphisms are transformations between topoi that preserve their logical structure (for the definition see \cite[p. 463]{Goldblatt}). One can speak on ``geometric logic'' as on that part of first order logic that is invariant under geometric morphisms (see \cite[p. 267]{Marquis}). The authors of \cite{tovariance} argue that since ``different topoi stand for different mathematical universes'', the tovariance principle well expresses the idea ``that physics should be as independent as possible of the mathematical framework'' and, consequently, it should be expressible in the invariant mathematical language. The idea is doubtlessly sound, but the postulate that the invariants in question should be those relative to geometric morphisms is controversial. The authors of the idea themselves have now serious reservations in this respect, ``since some of our constructions turned out to be non-geometric in this sense'', although they acknowledge that there are some possibilities to make these constructions geometric \cite[p. 493]{LandsmanFound}. The applicability of the tovariance principle (or its modifications) to the realm of infinitesimal objects is not that obvious at present. To solve this question would require a bit of constructive mathematics (e.g. to study some properties of Weil algebras spectra in topoi) and geometric logic to decide which properties of infinitesimals are expressible in terms of first order geometric logic and thus are preserved by geometric morphisms. 

In the present paper, we focus on ``gravitational effects'' within an ``infinitesimal neighbourhood`` such as a monad $\cM(x), \, x \in M$. Einstein's equivalence principle states that ``at every space-time point in an arbitrary gravitational field it is possible to choose a `locally inertial coordinate system' such that, within a sufficiently small region of the point in question, the laws of nature take the same form as in unaccelerated Cartesian coordinate systems in the absence of gravitation''\cite[p. 68]{Weinberg}. This formulation is usually called strong equivalence principle since it refers to all laws of nature (not only to gravitational laws). Its weak form establishes the local equivalence between gravity and acceleration: a gravitational field can locally be eliminated (or generated) by a suitable choice of reference frame (see,  for instance, \cite[p. 121-123]{Schutz}). It can locally be eliminated by the fact that at any point $x\in M$ of a (curved) space-time manifold $M$ there exists a flat tangent space $T_xM$, and any sufficiently small neighbourhood of $M$, containig $x$, sufficiently well approximates it. Physically, this can be implemented by free fall: freely falling particles can be thought of (locally) as moving in a flat tangent space-time. Let us also notice that the existence of the tangent space-time at each point of a given space-time manifold is a purely geometric effect, independent of any dynamical equations (like Einstein's equations). 

In this sense, the weak equivalence principle is a purely geometric effect, suitably interpreted in physical terms. In the present study, we go even deeper in identifying its geometric rooting. By changing to the topos $\cE $, containing the enriched real line $R$, we are able, so to speak, to  resolve the point $x$ of tangencey in $T_xM$ to obtain an insight into its infinitesimal structure. The Proposition of the preceding section gives the result of this procedure.

The fact that on $U \subset M$ we have a flat connection $\lambda $ corresponds to the Einstein weak equivalence principle: gravitational field can locally (on $U$) be elimnated. The Proposition says that this is so because the flat connection $\lambda $ on $U$ comes about from the framing $k_y: D(n) \to \cM(y)$, for evey $y \in U$. In other words, a ``synchronisation'' of local connections on every monad $\cM(y), \, y \in U$ gives a flat connection on $U$. However, the implication goes both ways: the flat connection coming about from the framing $k_y$ implies a flat connection on $U$. We are confronted with a strong interaction between a macroscopic neighbourhood $U$ and its constituent monads.

This is purely geometric effect (in the sense of SDG), independent of any dynamical equations, just as the local flatness of $M$ is a purely geometric effect (in the sense of the standard geomtry) coming from the existence of the flat tangent space to $M$ at a given its point. The question of whether the latter effect could be enhanced by some dynamical equations was once extensively discussed in relation to the so-called Mach's principle. Roughly speaking, the question was to which extent a global structure of space-time (e.g. global mass-energy distribution in space-time) can influence its local structure (e.g. its inertial properties); see, for instance \cite{BarbourPfister,Raine75}. Here, we leave this problem aside.

The weak equivalence principle asserts that on any local neighbourhood $U$ (understood macroscoplically) in a space-time manifold $M$ one can always choose a flat connection. Of course, there exist many other connections on $U$. The Proposition of the preceding section translates this situation into infinitesimal neighbourhoods (monads). It states that on every monad $\cM(y), \, y \in U_x$, the curvature vanishes if and only if it vanishes on the entire $U_x$. If, however, we consider a non-flat connection on monads in $U_x$, the components of the curvature tensor still assume only infinitesimal values on them \cite{Universe17}. On the other hand, if we consider a non-flat connection on $U_x\subset M$, the infinitesimal curvature contributions from monads in $U_x$ are integrated up to non-zero non-ifinitesimal values (e.g. \cite[p.235]{MoerRey}). In this way, the geometric seeds of classical gravity are hidden in infinitesimal neighbourhoods of spacetime manifold. If this is so, one should look for methods of quantising gravity in infinitesimal coordinates and ifinitesimal values of curvature. It looks as if nature had chosen to hide the infinitesimal regions of spacetime, on which the quantisation of gravity should be performed, from the direct  access of our reasoning. 

There is another field of applications of the above results, namely the theory of classical singularities in cosmology and astrophysics. As argued in \cite{Universe17}, if on ``infinitesimal neighbourhoods'' curvature vanishes (or is infinitesimal), perhaps curvature singularities could be avoided by changing to suitable categories.

The metric structure is crucial for general relativity, however, its status in quantising gravity is far from being clear. In our categorical approach on ``infinitesimal neighbourhoods'', the interpretation of metric in terms of a distance between points is meaningless. In spite of this, a metric can be defined on such neighbourhoods as a quadratic differential form, essentially with respect to points within the second order neighbourhoods, $x \sim_2 y$.

It is interesting that the metric, defined in this way, is a pseudo-Riemannian metric\footnote{Or rather semi-Remanian, but we prefer to stick to Kock's terminology.} (however, without the possibility of distinguishing space and time directions), and onlynly later, as a next step, it is possible  to assure its positiveness by imposing additional conditions. If this reflects a deeper regularity, rather than being a side-effect of our still imperfect understanding of the infinitesimal level, we could speculate that on this level the distinction between space and time is blurred, and we are confronted with a sort of a pregeometric entity rather than with a usual space-time.

The latter suggestion could be strengthened by taking into account the quasi-metric $\mathrm{dist}(x,y) \leq k$. Of course, one cannot associate with it any direct physical meaning (at least at the present state of our understanding), certainly not a usual distance. However, since $k$ in this formula denotes the ``$k$th order of an infinitesimal object'', it determimes a ``size'' of an object $D_k(n)$ (we should remember that $D_k(n) \subseteq D_l(n)$ if $k \leq l$). And this quasi-metric is ``quantised'' (discrete) since it has its values in $\NN $. We could borrow from D. Mumford the expression (used by him on quite different occasion, quoted after \cite[p. 82]{Kock09}) to call the dist-function a ``disembodied metric''.

Of course, all the above results are valid under the assumption that when dealing with  ``very small neighbourhoods'' we have to change from the SET environment to that of a suitable category in which synthetic differential geometry makes sense. It is a radical step since it enforces a weakening of standard logic. Such a step can be justified only by results it could lead to.

\appendix
\section{Appendix}

\subsection{Weil Algebras}
There are many equivalent ways of defining Weil algebra. Here we choose the one which allows for direct interpretations in categories. 

Let $R$ be a ring object in a category $\mathcal{E}$.  
A Weil algebra over $R$ is a commutative algebra $W$ such that its underlying $R$-module is of the form $R \oplus R^n$ with $(1, \underline{0})$ as a multiplicative unit; moreover, every elemnt $(0, \underline{v}), 
\, \underline{v} \in R^n $, is nilpotent.

We define the spectrum of a Weil algebra W as the object of $R$-algebra maps $W \to R$,
\[
\mathrm{Spec}_RW = R^W.
\]
Objects of the form $\mathrm{Spec}_RW$ are called (formal) infinitesimal objects.
There is the canonical map $\alpha : W \to R^{\mathrm{Spec}W}$. The generalisation of the Kock-Lawvere axiom states that $\alpha $ is an isomorphism.

\subsection{Formal Manifolds}
Here we give somewhat simplified definition of formal manifolds; for the full treatment see \cite[p. 68-75]{Kock2006}. First, let us define a crucial concept. A map $P \to Q$ is said to be an \'etale map if, for every infinitesimal object $K$ and any commutative diagram of the form
\begin{diagram}
1			&\rTo^{\underline{0}}    &K\\
\dTo  &                        &\dTo\\                   
P     &\rTo                    &Q
\end{diagram}
there is a unique diagonal fill-in arrow $K \to P$.

If $U \rightarrowtail R^n$ is a monic (``injective'') \'etale morphism, $U$ is said to be an $n$-dimensional model object. We now define an $n$-dimensional formal manifold as an object $M$ such that there exists a regular epic (``surjective'') morphism $\bigsqcup_i U_i \twoheadrightarrow M$, where each $U_i \rightarrowtail M$ is a monic \'etale morphism with each $U_i$ being an $n$-dimensional model object.
 
Of course, $\bigsqcup_i U_i \twoheadrightarrow M$ is a covering family of $M$ by jointly epic class of morphisms. Formal manifolds can be equipped with a topology. The concept of monad and that of ``$k$th neighbourhood of the diagonal'' can be defined for formal manifolds (see \cite[p. 68-75]{Kock2006}). One can also give a slightly weaker definition of a formal manifold for which the model objects are sets of the form $D_{\infty }(n)$ \cite{Kock80}.

\end{document}